\documentclass{aa}
\usepackage{graphicx,epsfig,color}
\usepackage{times,latexsym,txfonts,amssymb}
\usepackage{natbib}
\usepackage{supertabular,multirow,longtable,rotating}

\def\kms{km\,s$^{-1}$}
\def\Ebv{\ensuremath{{\mathrm E}_{\mathrm (B-V)}}}
\def\fH2{\ensuremath{f_{H_2}}}

\bibpunct{(}{)}{;}{a}{}{,}

\begin{document}

\title{Dense molecular clouds in the SN\,2008fp host galaxy~\thanks{%
This paper is based on observations made with ESO Telescopes at Paranal Observatory under program IDs
081.D-0558, 081.D-0697, and 082.D-0004}}

\author{N.\,L.\,J.~Cox \inst{1} and F.~Patat \inst{2}}

\institute{Instituut voor Sterrenkunde, KU~Leuven, Celestijnenlaan 200\,D, B-3001 Leuven, Belgium \\
          \email{nick.cox@ster.kuleuven.be}
          \and
          European Organization for Astronomical Research in the Southern Hemisphere (ESO), 
          Karl Schwarzschild-Str. 2, G-85748\\ Garching bei M\"unchen, Germany \\
          \email{fpatat@eso.org}
          }

\date{Accepted for publication in A\&A.}

\abstract{%
Supernovae (SNe) offer a unique opportunity to study physical properties, small-scale structure, 
and complex organic chemistry of the interstellar medium (ISM) in different galaxies.
}{
Interstellar absorption features, such as atomic and molecular lines as well as diffuse interstellar bands (DIBs), 
can be used to study the physical properties of extra-galactic diffuse interstellar clouds.
}{
We used optical high-resolution spectroscopy to study the properties of the ISM in the SN\,2008fp host galaxy, 
ESO428-G14. The properties of intervening dust were investigated via spectropolarimetry.
}{
The spectra of SN\,2008fp reveal a complex of diffuse atomic clouds at radial velocities 
in line with the systematic velocities of the host galaxy. In addition, 
a translucent ($A_\mathrm{V} \sim 1.5$~mag) cloud is detected at a heliocentric velocity 
of 1770~\kms\  (redshifted by 70~\kms\ with respect to the system velocity). This cold dense cloud 
is rich in dense atomic gas tracers, molecules, as well as diffuse interstellar bands. 
We have detected both C$_2$ and C$_3$ for the first time in a galaxy beyond the Local Group.
The CN (0,0) band-line ratios are consistent with an excitation temperature of T = 2.9 $\pm$ 0.4~K.
The interstellar polarisation law deviates significantly from what is observed in the Galaxy, 
indicating substantial differences in the host dust/size composition.
No variations over a period of about one month are observed in any of the interstellar medium tracers.
}{
The lack of variability in the extra-galactic absorption line profiles implies that the absorbing 
material is not circumstellar and thus not directly affected by the SN event.
It also shows that there are no significant density variation in the small-scale structure of the 
molecular cloud down to 100~AU. C$_2$ is used to probe the cold diffuse ISM density and temperature.
Here we also use observations of CN in a distant galaxies, though for now still in a limited way, 
for \emph{in situ} measurements of the cosmic background radiation temperature.
}


\titlerunning{Dense molecular clouds in the SN\,2008fp host galaxy}
\authorrunning{N.\,L.\,J.~Cox \& F.~Patat}

\maketitle

\section{Introduction}\label{sec:intro}

Multi-epoch high-resolution optical spectroscopy of nearby Type\,Ia supernovae have been presented in relation  to the study of \ion{Na}{i}
and \ion{Ca}{ii} to probe the circumstellar matter associated with the progenitors of these events. This has led to the discovery of
time variability in high-resolution \ion{Na}{i} absorption line systems for some supernovae (see e.g. \citealt{2007Sci...317..924P};
\citealt{2009ApJ...702.1157S}). Several similar events were detected using low-resolution spectroscopy (\citealt{2009ApJ...693..207B};
\citealt{2010AJ....140.2036S}). However, not all Type\,Ia SNe show this behaviour (e.g. \citealt{2013A&A...549A..62P}). Furthermore,
\citet{2011Sci...333..856S} found that for a large sample of Type\,Ia SNe, the absorbing material (\ion{Na}{i}) tends to be blue-shifted
with respect to the strongest sodium absorption line, which has been suggested as an indication of outflows from  the supernova
progenitor systems. This is expected for single-degenerate (SD) systems but is much less probable for double-degenerate (DD) systems.
This result needs to be taken with caution as the strongest sodium absorption component does not necessarily have the same
velocity as the disk gas at the position of the supernova (e.g. \citealt{2012ApJ...748L..11R}).
\citet{2013ApJ...779...38P} showed that some supernovae -- all of which were classified as blue-shifted -- display anomalously high 
\ion{Na}{i} column densities with respect to the amount of dust extinction, suggesting a relation to the outflowing circumstellar gas. 
Pathways for producing this enhancement were given for both the SD and DD scenarios.

The increasing availability of high-resolution high-sensitivity spectra (in particular when averaging over multiple epochs) has provided the
possibility to study the physical properties of the ISM in other galaxies in detail. Until recently, only a few supernovae, such as SN\,1987A, 
could be studied in detail. Now, bright stars in the Magellanic Clouds (\citealt{2006A&A...447..991C};
\citealt{2006ApJS..165..138W}) and the Andromeda and Triangulum galaxies (\citealt{2008A&A...480L..13C,2008A&A...492L...5C}) are
accessible - with some effort - to optical (high-resolution)  spectroscopy. To proceed in studying distant galaxies
it is becoming readily possible to use nearby SNe, similar to the exploitation of SN\,1987A to study the LMC (\citealt{1987ESOC...26..511G};
\citealt{1988PASAu...7..527P}), as background candles to probe the (physical) properties of the ISM such  as dust extinction
and UV field
strength in their host galaxy through the observation of atoms, molecules, and diffuse interstellar bands (DIBs). 
However, sufficiently bright SNe are required to obtain high-S/N spectra at high-resolution, which limits the number of possibilities 
for such studies. Furthermore, not all SNe will probe significant columns of interstellar matter, in particular if they are located 
in the approaching side of the respective galaxy. 

Although rare, an increasing number of SNe have been observed with intermediate to high spectral resolution to probe the atomic
and molecular diffuse ISM, including DIBs, in the disks and halos of galaxies beyond the Local Group, such as the Centaurus group 
(\citealt{1985ApJ...299..852D}; \citealt{1987AJ.....94..651R}; \citealt{1989A&A...215...21D}), the Virgo cluster 
(\citealt{1990AJ.....99.1476S}; \citealt{1991ApJ...383L..41M}) and the M81 group (\citealt{1994ApJ...420L..71B}).
The most detailed studies of both resolved molecular absorption lines and resolved DIB profiles arising from the cold diffuse 
ISM in galaxies beyond the Local Group were presented in \citet{2005A&A...429..559S} and \citet{2008A&A...485L...9C}. In addition, interstellar 
absorption features such as the UV bump and the strong 5780 DIB have also been detected in distant damped Ly$\alpha$ (DLA) systems 
(\citealt{2004ApJ...614..658J}; \citealt{2006ApJ...647L..29Y}; \citealt{2008MNRAS.383L..30E}, \citealt{2012MNRAS.tmp..159S}).

Here we present a detailed study of the properties of the ISM in the SN host galaxy and the Galactic halo as probed by
SN\,2008fp. The spectra were obtained as part of an ongoing campaign to observe atomic line variablity in bright Type\,Ia SNe
in an effort to distinguish between the SD and DD scenarios for Type\,Ia SNe (cf. \citealt{2013A&A...549A..62P}). In
Sect.~\ref{sec:obs} we present the observations and data reduction and give basic properties of the SN and its host galaxy.
The detected Galactic and extra-galactic interstellar absorption lines are presented in Sect.~\ref{sec:sn2008fp-ISM}.
In Sect.~\ref{sec:discussion} we discuss the molecular content, physical cloud conditions, interstellar line variability and the
presence and behaviour of extra-galactic molecules and DIBs in the SN host galaxy. We conclude with a summary in Sect.~\ref{sec:conclusion}.

\section{Observations}\label{sec:obs}

High-resolution UVES spectra were obtained in service mode for supernova 2008fp. SN\,2008fp is a normal Type\,Ia 
SN (\citealt{2008CBET.1509....1W}; \citealt{2011AJ....142..156S}) discovered by \citet{2008CBET.1506....1P} in the 
galaxy \object{ESO428-G14}, which is at $\sim$26~Mpc. The reported galaxy type is SABr(0)-pec 
(\citealt{1991RC3.9.C...0000d}) and Seifert 2. Its heliocentric system velocity is 1698$\pm$20~\kms\
(\citealt{2003AJ....126.2268W}). SN\,2008fp is located in the sky at a Galactic longitude of 241.9\degr\ and at a 
Galactic latitude of -8.0\degr. \citet{2013ApJ...779...38P} derived $A_\mathrm{V} = 0.71_{-0.08}^{+0.10}$~mag and
$R_\mathrm{V} = 1.20_{-0.14}^{+0.26}$ from a light-curve analysis.

Spectra were obtained at different phases of the SN evolution. SN\,2008fp was observed at 6, 11, 17, and 39~days
after the reported discovery. These correspond to days -2, +3, +9, and +31 with respect to the optical maximum
light attained on MJD $\approx$ 54728 (\citealt{2011AJ....142..156S}). Thus, the set of spectra covers
approximately one month of temporal evolution. Table~\ref{tb:observations} gives a summary of the observations.
The data were processed with the standard UVES pipeline routines. All individual spectra were corrected for the
heliocentric velocity before averaging of individual frames at each epoch. In addition, we noticed a small shift
(of about 1 pixel) in peak velocities of the individual cloud components between  spectra taken at different
epochs. Further analysis of the sky emission lines (at 5889.95, 5895.924, and 6304~\AA) as well as the Galactic
foreground absorption lines showed a  small deviation in wavelength calibration between exposures taken with the
red arm. The spectra taken with the blue arm  did not suffer from this small shift. All individual spectra were
corrected for this effect. The sky emission lines and telluric absorption lines were measured to confirm the
obtained spectral resolution of 6~\kms. The achieved signal-to-noise ratio (S/N) in the averaged spectra is about 
60 to 80 for the blue range (e.g. for CN, \ion{Ca}{i}, CH, CH$^+$), while for the red range it is about 90 to 120 
(e.g. 5797, 5780 and 6613~\AA\ DIBs). In the near-infrared it is again lower; the S/N is about 50 -- 60 for the 
C$_2$ transitions.

\begin{table}
\caption{Observation log for SN\,2008fp.}
\label{tb:observations}
\centering
\resizebox{\columnwidth}{!}{
\begin{tabular}{lllll}\hline\hline
Target          & Date          & Epoch  & UVES setting\tablefootmark{a}& Exposure  \\ 
                & (yyyy-mm-dd)  & (day)  &             & time (s)  \\ \hline
SN\,2008fp      & 2008-09-17    & $+6$   & Blue        & 7\,200    \\
                &               &        & Red1        & 3\,600    \\
                &               &        & Red2        & 3\,600    \\
                & 2008-09-22    & $+11$  & Blue        & 7\,200    \\
                &               &        & Red1        & 3\,600    \\
                &               &        & Red2        & 3\,600    \\
                & 2008-09-28    & $+17$  & Blue, Red1  & 3\,600    \\
                & 2008-10-20    & $+39$  & Blue, Red1  & 7\,200    \\ \hline
\end{tabular}
}
\tablefoot{
\tablefoottext{a}{
UVES spectral coverage (approximately) for each setting: blue [3300 -- 4550~\AA], 
red1 [4600 -- 5600~\AA\ + 5700 -- 6650~\AA], red2 [6700 -- 8500~\AA\ + 8650 -- 10500~\AA].
The observed epoch is calculated as the number of days since 2008-09-11, the
date of discovery.}}
\end{table}

\section{Interstellar medium towards SN\,2008fp}\label{sec:sn2008fp-ISM}

First inspection of the UVES spectra of SN\,2008fp revealed immediately that the line-of-sight probes a rich host-galaxy ISM.
Atoms, molecules, and DIBs are detected at radial velocities corresponding to that of the Milky Way and the host galaxy, 
ESO\,428-G14. Next we present the observed interstellar lines in the Milky Way and SN host galaxy.
The Galactic and extra-galactic \ion{Ca}{ii} and \ion{Na}{i} lines are discussed in more depth in Cox et al. (in preparation), 
which will present an extensive analysis of both Galactic and extra-galactic sodium and calcium doublets for a larger sample of SNe.

\subsection{Galactic foreground ISM}

In addition to the strong Galactic \ion{Ca}{ii} and \ion{Na}{i} components present between 0 and 120~\kms, we also detected CH$^+$ and
several DIBs. Equivalent widths and radial velocities are given in Table~\ref{tb:sn2008fp-MW}. Typical radial velocities are
between 50 and 70~\kms.  Using the DIB strength versus interstellar reddening relation for field stars derived by
\citet{2008A&A...480..133L}, the 5797 and 6613~\AA\ DIB strengths give a Galactic foreground reddening of $\sim$0.15 -- 0.20~mag.
This is in excellent agreement with the Galactic foreground reddening, \Ebv\ = 0.2~mag, estimated from the all-sky dust extinction
map (\citealt{1998ApJ...500..525S}). The total column densities for Galactic \ion{Ca}{ii} is $1.4 \times 10^{13}$~cm$^{-2}$.

\begin{table}[ht!]
\caption{Galactic CH$^+$ and diffuse bands in the lines-of-sight towards SN\,2008fp.}
\label{tb:sn2008fp-MW}
\centering
\begin{tabular}{lll}\hline\hline
Line            &  EW (m\AA)\tablefootmark{a}    &$v$ (\kms)\\ \hline
CH$^+$          &   5 $\pm$ 2    & 67.4     \\ 
5780.55 DIB     & 115 $\pm$ 10   & 59.6     \\ 
5797.08 DIB     &  25 $\pm$ 3    & 68.3     \\ 
5849.81 DIB     &  19 $\pm$ 2    & 50.7     \\ 
6613.63 DIB     & 38 $\pm$ 7     & 62.1     \\ 
\hline
\end{tabular}
\tablefoot{
\tablefoottext{a}{Equivalent width uncertainties have been computed following the prescription in Vos et al (2011).}
}
\end{table}

\begin{table}[th!]
\caption{Atomic, molecular, and diffuse bands detected in the
line-of-sight toward SN\,2008fp.}
\label{tb:sn2008fp_ISlines}
\centering
\resizebox{\columnwidth}{!}{
\begin{tabular}{llrl}\hline\hline
line                            & $v_{\rm helio}$\tablefootmark{a} & \multicolumn{1}{c}{EW}    & log\,N\tablefootmark{a}   \\
                                & (\kms)                & \multicolumn{1}{c}{(m\AA)}           & (cm$^{-2}$)                 \\ \hline
\ion{Na}{i}\,UV (doublet)       & 1771.5 $\pm$ 0.7      & 128 $\pm$ 26     & 14.43 $\pm$ 0.03                 \\
                                &                       & 67 $\pm$ 13      &                            \\
\ion{Ca}{ii}\,HK (doublet)      & 1700 -- 1780          &                  & 13.39 $\pm$ 0.02                \\
\ion{Fe}{i} (3719.9\AA)         & 1770.4, 1784.4        & 44 $\pm$ 5       & 12.96 $\pm$ 0.03           \\
\ion{Fe}{i} (3859.9\AA)         &                       & 26 $\pm$ 6       &                            \\
\ion{Ca}{i}                     & 1770.5, 1780.7        & 60 $\pm$ 5       & 11.35 $\pm$ 0.02        \\
\ion{Na}{i}\,D (doublet)        & 1726 -- 1790          &                  & 14.35 $\pm$ 0.06                \\
\ion{K}{i} (doublet)            & 1770 -- 1791          &                  & 12.03 $\pm$ 0.03                \\
\ion{Ti}{ii}                    & 1775                  & 46 $\pm$ 9       & 12.1  $\pm$ 0.1            \\
                                &                       &                  &                         \\
CN B-X (0-0) R$_0$              & 1767.8                & 84 $\pm$ 3       & 13.40 $\pm$ 0.03  \\
CN B-X (0-0) R$_1$              & 1768.7                & 37 $\pm$ 3       & 13.12 $\pm$ 0.06  \\
CN B-X (0-0) P$_1$              & 1768.3                & 19 $\pm$ 2       &                         \\
CN B-X (1-0) R$_0$              & 1769.3                & 18 $\pm$ 4       & 13.7  $\pm$ 0.1         \\
CH$^+$ A-X(0-0) R(0)            & 1773.2, 1784.3        & 16 $\pm$ 3       & 13.23 $\pm$ 0.22           \\
CH$^+$ A-X(1-0) R(0)            &                       &  8 $\pm$ 2       &                            \\ 
CH A-X (0-0)                    & 1768.9, 1779.2        & 27 $\pm$ 4       & 13.45 $\pm$ 0.36                \\
CH B-X (0-0)                    &                       & 11 $\pm$ 3       &                         \\
C$_2$ A-X (2-0) R0              & 1766.8                & 16 $\pm$ 2       & 13.23 $\pm$ 0.06\tablefootmark{b}       \\
C$_2$ A-X (2-0) R2              & 1768.9                & 19 $\pm$ 2       & 13.70 $\pm$ 0.05\tablefootmark{b}       \\
C$_2$ A-X (2-0) R4-R8           & 1768.1                & 15 $\pm$ 2       & 13.68 $\pm$ 0.06\tablefootmark{b}       \\
C$_2$ A-X (2-0) R6              & 1769.5                & 8  $\pm$ 3       & 13.44 $\pm$ 0.20\tablefootmark{b}       \\
C$_2$ A-X (2-0) Q2              & 1765.9                & 25 $\pm$ 3       & 13.72 $\pm$ 0.06\tablefootmark{b}       \\
C$_2$ A-X (2-0) Q4              & 1770.2                & 21 $\pm$ 3       & 13.64 $\pm$ 0.04\tablefootmark{b}       \\
C$_2$ A-X (2-0) Q6              & 1772.6                & 15 $\pm$ 3       & 13.50 $\pm$ 0.10\tablefootmark{b}       \\
                                &                       &                  & $\Sigma(N) =$                   \\
                                &                       &                  & 14.29 $\pm$ 0.07\tablefootmark{b}\\
C$_3$ \~A-\~X (Q-branch)        & 1768.0                & 20 $\pm$ 5       & 13.23 $\pm$ 0.15\tablefootmark{c} \\
                                &                       &                  &                         \\
5780.55~DIB                     & 1787 $\pm$ 10         & 81 $\pm$ 5       &                                 \\
5797.08~DIB                     & 1780 $\pm$ 5          & 60 $\pm$ 10      &                                 \\
5849.81~DIB                     & 1778 $\pm$ 5          & 34 $\pm$ 5       &                                 \\
6196.00~DIB                     & 1776 $\pm$ 14         & 29 $\pm$ 6       &                                 \\
6283.85~DIB                     & 1796 $\pm$ 15         & 180$\pm$ 30      &                               \\ \hline
\end{tabular}
}
\tablefoot{
\tablefoottext{a}{Column densities and radial velocities for the atomic species CH, CH$^+$, and CN are obtained from multiple-component 
profile fitting using VPFIT. Whenever possible, multiple transitions (e.g. doublets) of a single species were fitted simultaneously.
For \ion{Fe}{i}, CH$^+$, and CH, two velocity components (see table) were required to achieve the best fit.}
\tablefoottext{b}{Details for N(C$_2$) in Sect.~\ref{subsec:molecules}.}
\tablefoottext{c}{N(C$_3$) is derived following the procedure of \citet{2003ApJ...582..823O} (Sect.~\ref{subsec:molecules}).}
}
\end{table}

\begin{figure}[th!]
\centering
\includegraphics[width=\columnwidth,clip]{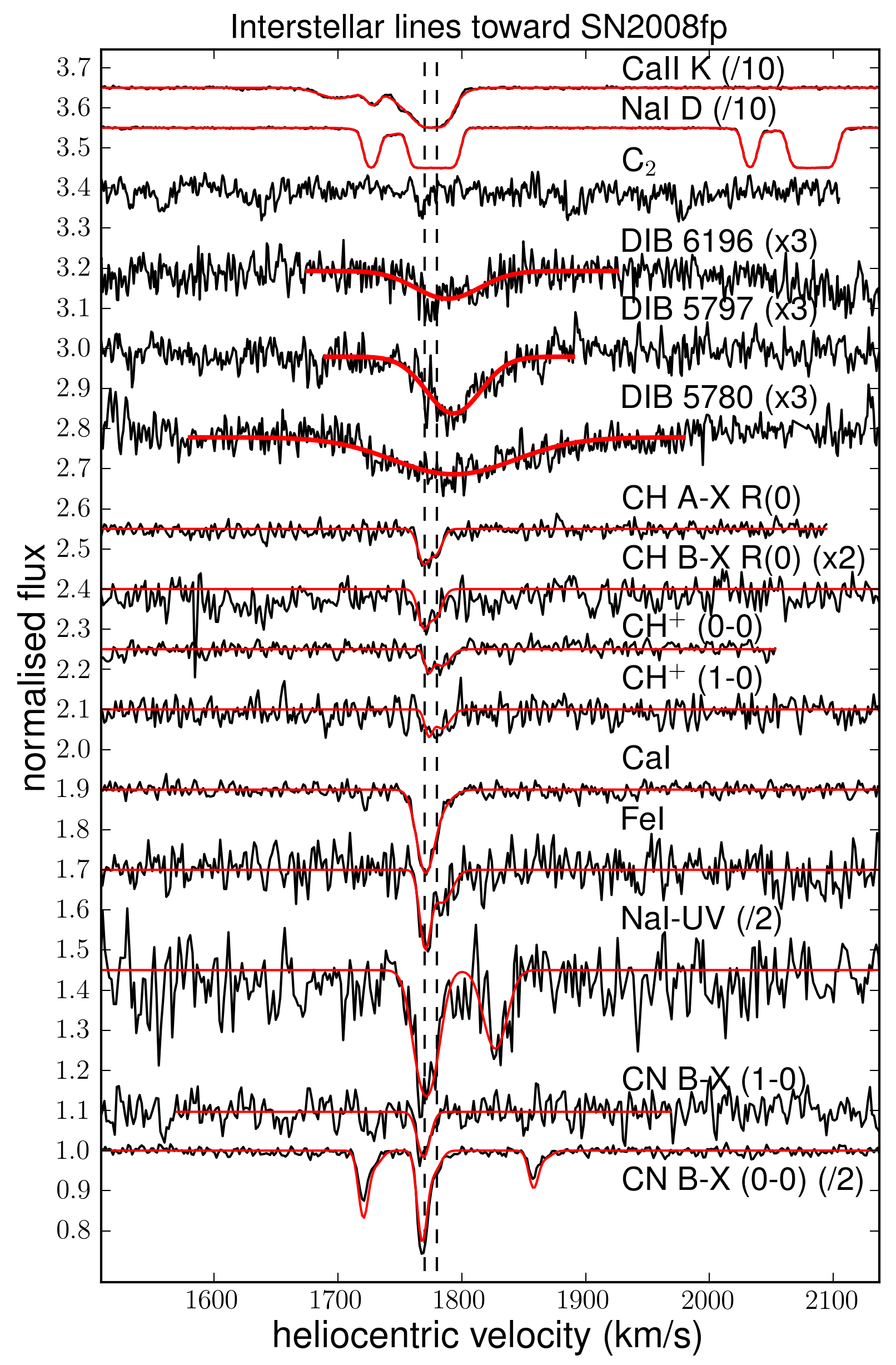}
\caption{Interstellar absorption lines towards SN\,2008fp. Synthetic line profiles obtained with VPFIT are 
shown as solid red curves overplotted on the observed spectra. Details of the models are given in Table~\ref{tb:sn2008fp_ISlines}.
The vertical dashed line are drawn at 1770 and 1778~\kms, the average peak radial velocity for the strongest interstellar components. The 5797 DIB and CH$^+$ appear slightly red-shifted with respect to the average radial velocity.
The shift for 5797 DIB corresponds to roughly 0.4~\AA\ or 20~\kms. 
Two velocity components separated by $\sim$8~\kms\ are discerned in the absorption profiles of CN, CH, and CH$^+$.}
\label{fig:sn2008fp_ISlines}
\end{figure}

\subsection{Supernova host galaxy ISM}

Both \ion{Na}{i}\,D and \ion{Ca}{ii}\,HK are detected at radial velocities corresponding to those of the host galaxy 
(Fig.~\ref{fig:sn2008fp_ISlines}). The main absorption complex is between +50 to +100~\kms\ with respect to the system velocity
at 1698~\kms. In addition, many other atomic and molecular species are detected at velocities corresponding to those of the main
components of \ion{Na}{i,} as shown clearly in Fig.~\ref{fig:sn2008fp_ISlines}. 
We used VPFIT~\footnote{Carswell \& Webb; \mbox{http://www.ast.cam.ac.uk/\\~rfc/vpfit.html}} to obtain densities for the atomic 
and molecular species. Whenever possible, multiple transitions of a particular species were used to correct for line 
saturation and improve the constraint of the fit.
The results for the the central radial velocities and the column density or equivalent width for the atomic and molecular species 
and DIBs, respectively, are listed in Table~\ref{tb:sn2008fp_ISlines}.
A detailed discussion of each of the atomic species, molecules, and DIBs in the SN\,2008fp host galaxy are given 
in the subsequent subsections.

\subsubsection{Extra-galactic atomic species}\label{subsec:atoms}

The strongest absorption features near the systemic velocity of the host galaxy are from \ion{Na}{i} and \ion{Ca}{ii}.
These are shown in the top two traces of Fig.~\ref{fig:sn2008fp_ISlines}. For both species the transitions between 1750
and 1800~\kms\ are saturated. Two weaker components are present at 1700 and 1730~\kms. 
In addition to these strong absorption systems the weaker transitions of \ion{Ca}{i}, \ion{K}{i}, and \ion{Na}{i}\,UV
are also detected around 1770--1780~\kms~(Fig.~\ref{fig:sn2008fp_ISlines} and Table~\ref{tb:sn2008fp_ISlines}). 
\ion{Ti}{ii} is only tentatively detected at a radial velocity of 1775~\kms\ with an equivalent width of 46~$\pm$~9~m\AA.
Other interstellar atomic species, such as \ion{Li}{i} and \ion{K}{i}\,UV, are not detected.
Total column densities derived with VPFIT are listed in Table~\ref{tb:sn2008fp_ISlines}. Whenever possible, multiple transitions, 
such as those for doublets, were included in the VPFIT fitting procedure to take saturation into account 
and thus improve the robustness of the total column densities.
The uncertainties in equivalent width were derived from the continuum signal-to-noise ratio, while uncertainties on the
column densities - except for C$_2$ and C$_3$-- are those reported by VPFIT. This results in differences in 
relative uncertainties for equivalent widths and column densities.

\subsubsection{Extra-galactic molecular lines and bands}\label{subsec:molecules}      

\paragraph{CH} The strong CH A-X (0-0) and B-X (0-0) absorption lines at 4300.313~\AA\ 
and 3886.4132~\AA, respectivley, are clearly detected towards SN\,2008fp (Fig.~\ref{fig:sn2008fp_ISlines}). 
Two velocity components separated by $\sim$8~\kms\ are present.
To take into account mild saturation of the strongest line, both lines are fitted simultaneously with VPFIT, 
which gives N(CH) $= 2.8^{+3.7}_{-1.6} \times 10^{13}$~cm$^{-2}$.

\paragraph{CH$^+$} Both the CH$^{+}$ A-X (0-0) R(0) and A-X (1-0) R(0) transitions at 4232.548 and 3957.70~\AA,\ respectively, 
are detected (Fig.~\ref{fig:sn2008fp_ISlines}). 
Two velocity components separated by $\sim$8~\kms\ are present.
A simultaneous fit to the two transitions gives N(CH$^+$) = $1.7^{+1.1}_{-0.7} \times 10^{13}$~cm$^{-2}$.

\begin{figure}[t!]
\centering
\includegraphics[angle=0,width=\columnwidth,clip]{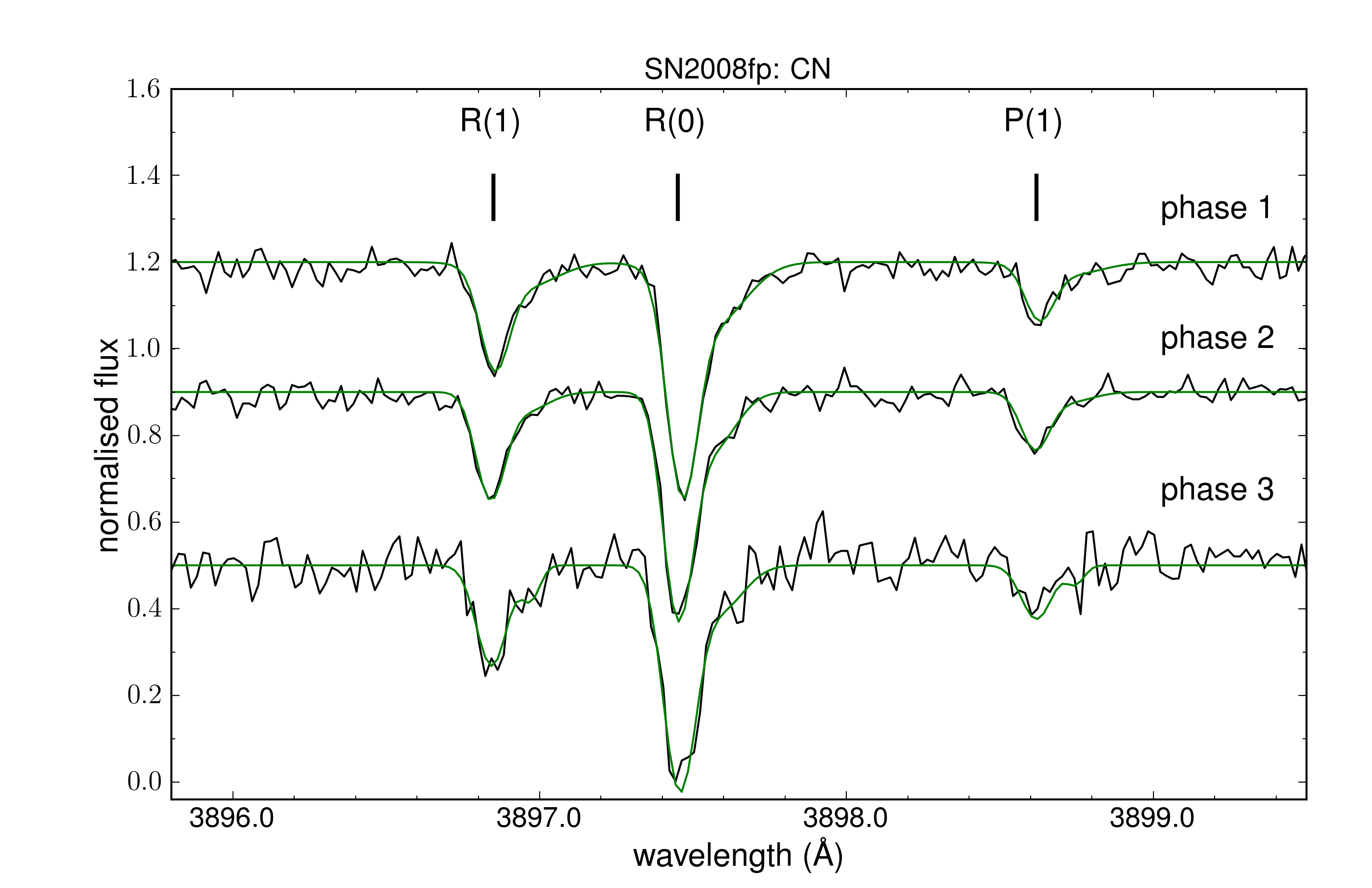}
\caption{CN R-X (0,0) spectrum observed toward SN\,2008fp at first 3 epochs. The indicated line transitions are plotted at a common radial velocity of 1770~\kms,
corresponding to the velocity of atomic and di-atomic lines associated to the host galaxy.
The VPFIT model is overplotted as a solid green curve.}
\label{fig:sn2008fp_cn}
\end{figure}

\paragraph{CN} The R$_0$, R$_1$, and P$_1$ lines of the CN\,(0,0) $B^2 \Sigma^+ - X^2 \Sigma^+ (0,0)$ vibrational band, 
as well as the R$_0$ line of the CN\, B-X (1-0) band at 3579.963~\AA,\ 
are detected towards SN\,2008fp (Figs.~\ref{fig:sn2008fp_ISlines} and~\ref{fig:sn2008fp_cn}, and Table~\ref{tb:CN}). 
The R$_0$ and R$_1$ line profiles have a weak red wing, suggesting
that the absorption system consists of at least two velocity 
components separated by $\sim$~9~\kms\ (similar to the velocity structure seen in CH and CH$^+$ line profiles). 
The R$_0$ profile peaks at a radial velocity of $\sim$1768~\kms. 

From the VPFIT line fitting we derive a total column density N(CN\,B-X(0-0)) = $2.6^{+0.9}_{-0.6} \times 10^{13}$~cm$^{-2}$. 
The Doppler width is $b = 3.7 \pm 1.6$~\kms\ for the main component. The second weaker component appears broader, but is poorly constrained.
This is similar to the curve-of-growth-corrected column density, N(CN B-X (0,0)) $\approx~2.5 \times 10^{13}$~cm$^{-2}$ (Sect.~\ref{sec:CN}).
We find a higher column density for the weaker line, CN\,B-X(1-0) line of N(CN) = $(5.0 \pm 1.1) \times 10^{13}$~cm$^{-2}$ 
(or log\,N(CN) = 13.7; optical thin limit, with $f = 0.0030$). 
The ratio N(CN(J=0)/N(CN(J=1) = 1.9, indicating the CN(J=0) line is weakly saturated (c.f. Sect.~\ref{sec:CN}).

\begin{figure}[t!]
\centering
\includegraphics[angle=0,width=\columnwidth]{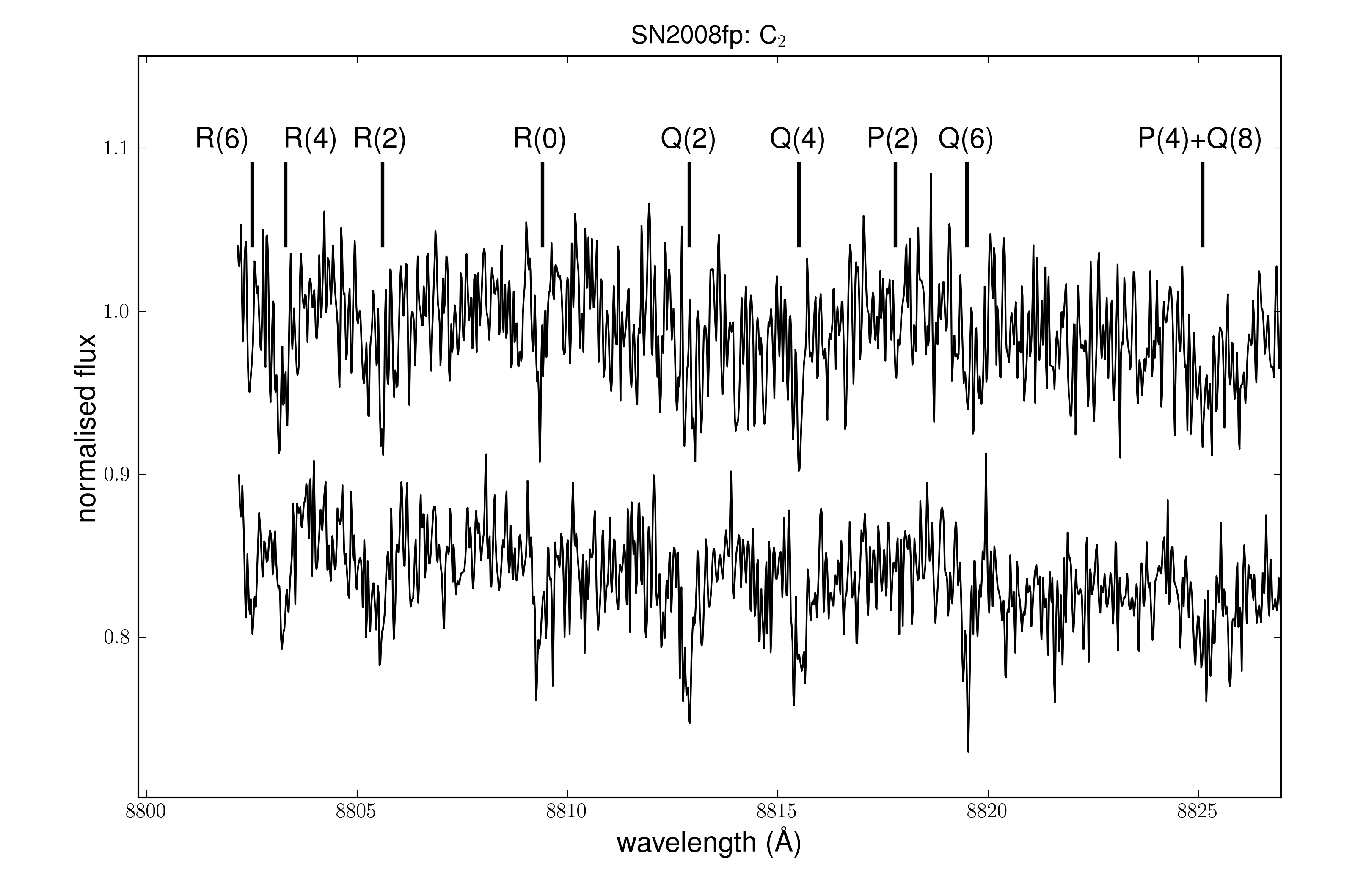}
\vspace*{-5mm}
\caption{C$_2$ spectrum observed toward SN\,2008fp. The indicated line transitions are plotted at a common radial velocity of 1770~\kms,
corresponding to the velocity of atomic and di-atomic lines associated to the host galaxy.}
\label{fig:sn2008fp_c2}
\end{figure}

\paragraph{C$_2$} The transitions of the C$_2$ (2,0) band detected at velocities of
the SN\,2008fp host galaxy are only shown for epochs 1 and 2 because only these are included in the UVES Red2 setting (Fig.~\ref{fig:sn2008fp_c2}). 
The indicated line transitions are plotted at a common radial velocity of 1770~\kms. 
A rotational diagram analysis of the level population of C$_2$\footnote{The web-tool (http://dib.uiuc.edu/c2/calc.html)
provided by B.J.~McCall was used to infer 
gas density and temperature from the measured C$_2$ column densities and to correct the observed column
density (J=0-6) to obtain the total (J$\leq$20) column density.} (see also \citealt{2007ApJS..168...58S}) 
shows that the interstellar cloud probed by SN\,2008fp has particle density 
$n_\mathrm{collisions} \approx$ 250~cm$^{-3}$ and rotational temperature $T_{\rm rot} \approx$~30~K.

\begin{figure}[t!]
\centering
\vspace{-5mm}
\includegraphics[angle=0,width=\columnwidth]{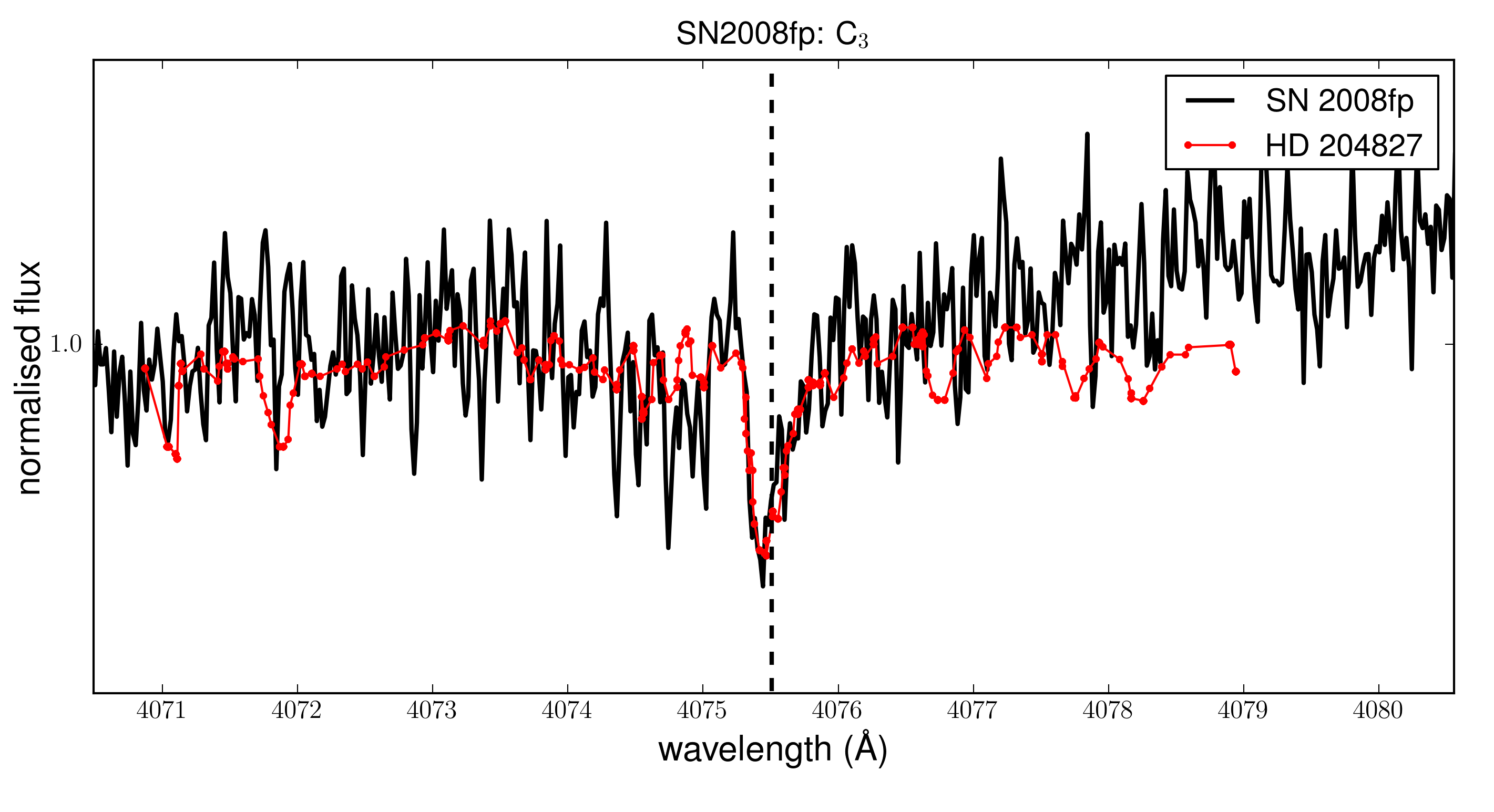}
\caption{{C$_3$ spectrum (black) observed toward SN\,2008fp. The vertical dashed line indicates a radial velocity of 1768~\kms.
The normalised C$_3$ spectrum of HD 204827 (\citealt{2003ApJ...595..235A}), shifted by 1768~\kms, is shown for comparison. }
}
\label{fig:sn2008fp_c3}
\end{figure}

\paragraph{C$_3$} The C$_3$ $Q$-branch ($\lambda_\mathrm{rest}$ = 4051.6~\AA; \citealt{2001ApJ...553..267M})
is detected at a radial velocity of 1768.0~\kms\, with an equivalent width of 20$\pm$5~m\AA. 
The observed C$_3$ spectrum is shown in Fig.~\ref{fig:sn2008fp_c3} together with the C$_3$ spectrum of HD\,204827 
(T$_\mathrm{C_3} \sim 42~K$; \citealt{2003ApJ...595..235A}).
Note the striking resemblance in profile shape and strength of the $Q$-branch peak at 4051.6~\AA\ in the two sightlines. 
From detailed modeling, \citet{2003ApJ...595..235A} found that for high gas densities both C$_2$ and C$_3$ excitation profiles are 
determined primarily by the kinetic temperature. In other words, the high-J levels of C$_3$ are not significantly populated in dense 
environments. 
Indeed, the low population of $R$-band levels, including the clear absence of the 4049.8~\AA\ $R$-band head, is indicative of a thermal
distribution at $\sim$50~K (\citealt{2003ApJ...595..235A}). 
In this case, the approximation that the observed Q-branch accounts for about half the total intensity of the C$_3$ band is valid 
(\citealt{2003ApJ...582..823O}). 
Thus, assuming an optically thin line, N(C$_3$) = (17 $\pm$ 5) $\times$ 10$^{12}$~cm$^{-2}$, similar to 
N(C$_3$) = (11.51 $\pm$ 0.87) $\times$ 10$^{12}$~cm$^{-2}$ derived for HD\,204827 by \citet{2003ApJ...595..235A}.
This result adds to the recent detection of extra-galactic C$_3$ in the SMC towards the peculiar line-of-sight of Sk\,143  
(\citealt{2013MNRAS.428.1107W}) with reported N(C$_3$) = (5.7 $\pm$ 0.6) $\times$ 10$^{12}$~cm$^{-3}$.

\paragraph{DIBs} The diffuse bands at 5780, 5797, 5849, 6196, and 6283~\AA\ are detected at radial velocities corresponding to the atomic
and molecular species reported above. 
We note that because of the radial velocity shift of the host galaxy ISM, the strong DIB at 6613~\AA\
is expected to be situated just outside the observed spectral range.
Radial velocities and equivalent widths are given in Table~\ref{tb:sn2008fp_ISlines}.

\subsubsection{Extra-galactic dust: interstellar polarisation}\label{results:pol}

Additional and independent information on the nature of the intervening material can be derived by analysing the continuum interstellar linear
polarisation (hereafter ISP) produced by dust along the line of sight (\citealt{1975ApJ...196..261S}). 
From the previous section it is clear that SN~2008fp suffers from substantial visual extinction. 
It is therefore natural to expect a marked interstellar polarisation due to (interstellar) dust, as in the case of SN\,2006X \citep{2009A&A...508..229P}.

\begin{figure}[t!]
\centering
\includegraphics[width=1.\columnwidth]{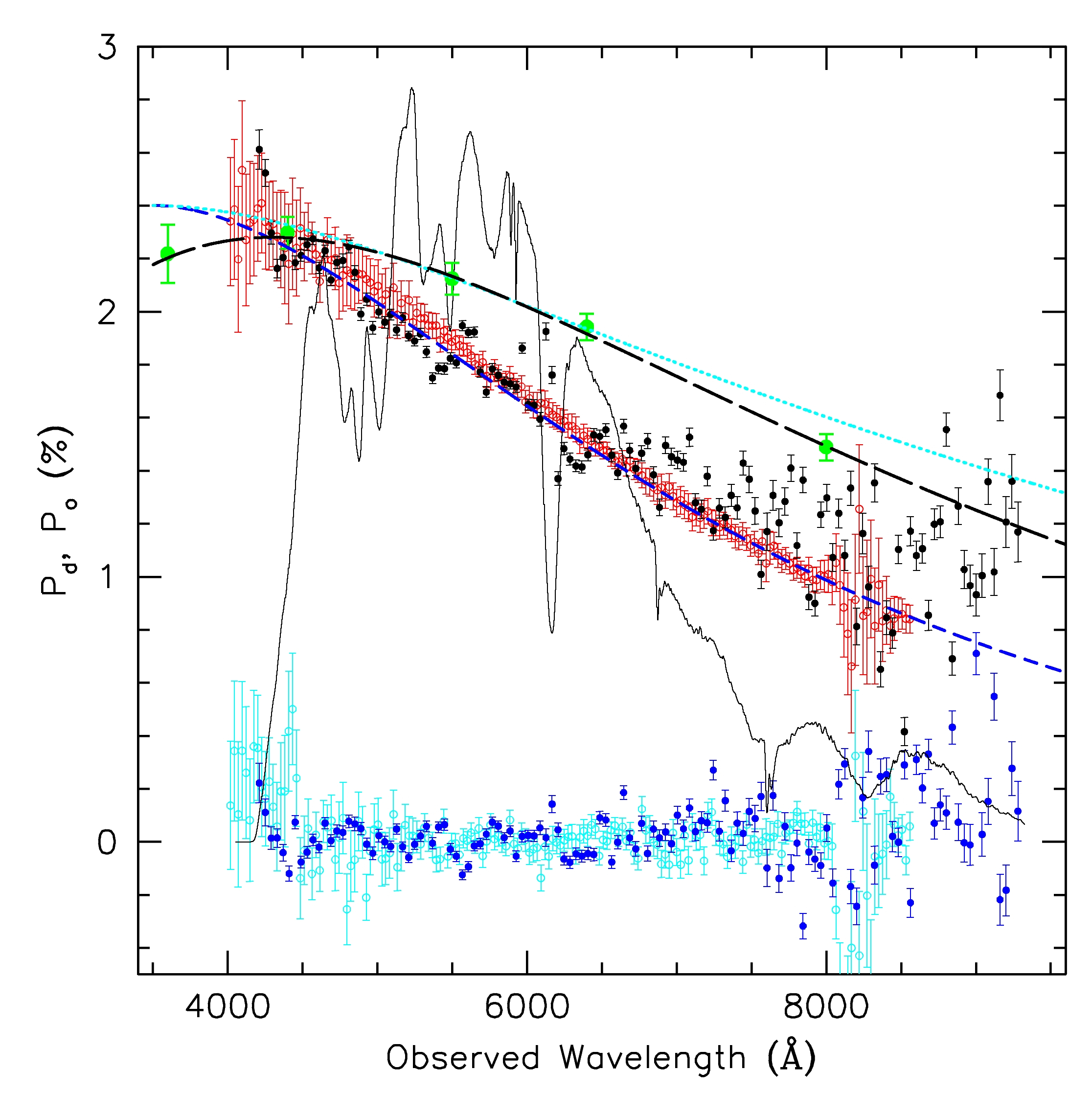}
\caption{Spectropolarimetry of SN\,2008fp along the dominant (upper) and orthogonal (lower) axes (filled symbols) on day~+6.  The
underlying solid curve traces the unbinned observed spectrum of SN~2008fp on the same epoch. For comparison the scaled data of SN\,2006X are
presented as well (empty symbols; \citealt{2009A&A...508..229P}). The short-dashed line is a best fit to the SN\,2006X data using a Serkowski law
($\lambda_{max}$ = 3500~\AA, $K$ = 1.30), while the dotted line is the same but with $K$ given by the \citet{1992ApJ...386..562W} relation
($K$ = 0.59).  The large, filled dots indicate the scaled broad-band polarimetry measurements of SN\,1986G (\citealt{1987MNRAS.227P...1H}). 
The long-dashed curve traces a best-fit Serkowski law ($\lambda_{max}$ = 4300~\AA, $K$ = 1.10).}
\label{fig:isp}
\end{figure}

We obtained spectropolarimetry of SN~2008fp on days $-$5, 0, +2 and +6 days from B maximum light. The data were obtained using FORS1 at the 
ESO-VLT and were reduced and analysed as described in \citet{2009A&A...508..229P}. Details on the observations will be given in a separate paper, 
which will also present a detailed study of the intrinsic polarisation properties of SN\,2008fp. 
Here we focus on the ISP in connection with the ISM properties we derived from high-resolution spectroscopy.
The data show a continuum polarisation that steadily  grows from the red to the blue, exceeding 2.2\% at 4200~\AA~(Fig.~\ref{fig:isp}). 
The polarisation wavelength dependency shows no signs of the maximum typically seen in highly reddened Galactic stars, which
occurs
on average around 5400 \AA\ (\citealt{1975ApJ...196..261S,1992ApJ...386..562W}).
The continuum polarisation angle measured on the combined  spectrum is 148.4$\pm$0.15 degrees. 
This is very well aligned to the local spiral  structure, as in other well-observed SNe and consistent with the orientation of dust grains 
along the Galactic magnetic field (\citealt{1987MNRAS.224..299S}).

However, the derivation of ISP in SN\,2008fp is made somewhat uncertain by two problems: 
i) the ISP appears to be more than a factor four smaller than in the extreme case of 2006X (\citealt{2009A&A...508..229P}); 
ii) no late-time polarimetric data are available. This implies that the SN imprints are expected to be more marked in 2008fp than in 2006X. 
Furthermore, the data suffer from a marked fringing redwards of 6000~\AA, increasing the noise level well above the photon noise
statistics in that spectral region.
This is illustrated in Fig.~\ref{fig:isp}, where we plot the decomposition of the observed Stokes parameters $Q$ and $U$ along 
the dominant (identified by the global polarisation angle) and the orthogonal axis, $P_d$ and $P_O$, respectively. 
To minimise the SN contribution to the observed polarisation we chose to use the last available epoch (day +6). 
As anticipated, SN\,2008fp displays a pronounced continuum polarisation along the dominant direction (148.4 degrees), 
on top of which some intrinsic features are detected at the positions of the main absorptions of the flux spectrum, 
chiefly Si~II 6355~\AA\ and Ca~II NIR triplet. 
While signatures of these features are also visible in the orthogonal component, the continuum polarisation is consistent with zero in
the wavelength range 4000-7500~\AA. At redder wavelengths there may be an additional continuum contribution, although this region is 
affected both by the Ca~II triplet and by fringing, most likely responsible for the observed scattering. 
The ``residual'' continuum polarisation is below the 0.2\%-level, which is typical of normal Type Ia (\citealt{2009A&A...508..229P}). 
The comparison between polarisation curves of SN\,2008fp and SN\,2006X is discussed in Sect.~\ref{sec:2006X}.

\section{Discussion}\label{sec:discussion}

In this section we discuss the properties of the extra-galactic ISM in the line-of-sight of SN\,2008fp.

\subsection{Extra-galactic diffuse interstellar bands}\label{subsec:DIBs}

The 5780~\AA\ DIB correlates with \Ebv\ in the MW, MCs, and DLA systems. Thus, DIB formation/survival and high dust content are
closely linked. Several surveys have found that the 5780~\AA\ DIB is,  for the same \ion{H}{i} column density, typically a factor
of 8 and 20 weaker in the LMC and SMC than in the Milky Way (\citealt{2006A&A...447..991C}; \citealt{2006ApJS..165..138W}). 
Moreover, the 5780~\AA\ DIB is three to six times weaker in DLAs for a given N(\ion{H}{i}) than in the MW sightlines, whereas the 6283 
is 4-10 times under-abundant (\citet{2006ApJ...647L..29Y}; \citealt{2008AJ....136..994L}). The observed 5780~\AA~DIB strength is 
consistent with the Galactic EW-\Ebv\ relation.

The equivalent widths of the 5780, 5797, 5849, and 6196~\AA\ DIBs predict, assuming a Galactic trend (\citet{2008A&A...480..133L}), 
an \Ebv\ between 0.20 and 0.55~mag. 
At low reddening values the scatter on these relationships can be as much as a factor of two.

\begin{figure}[t!]
\centering
\includegraphics[angle=0,width=\columnwidth,clip]{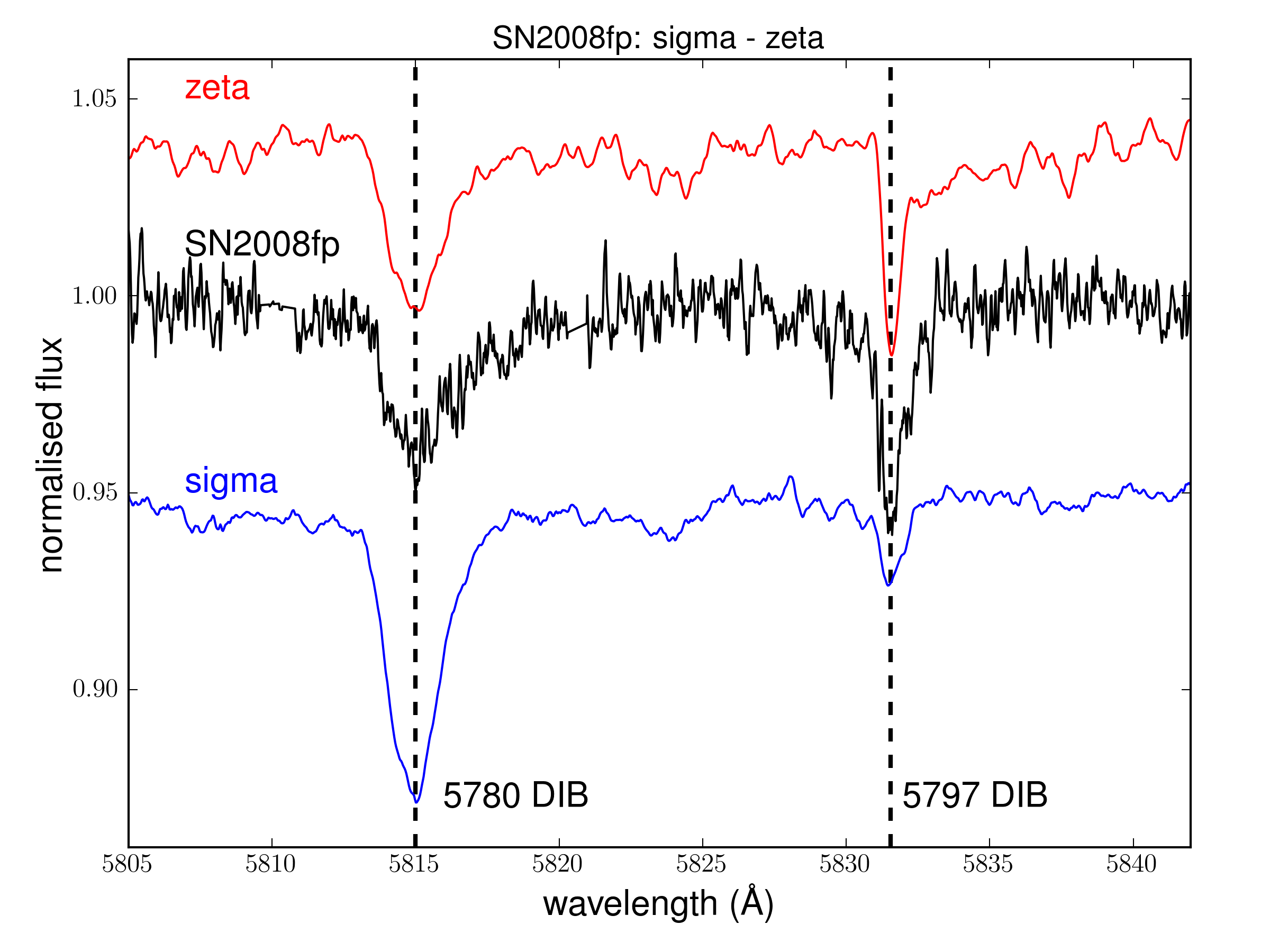}
\caption{Extra-galactic 5780 and 5797 DIBs at 1800 and 1789~\kms\ towards SN\,2008fp.
Typical $\sigma$ (HD\,144217) and $\zeta$ (HD\,149757) Galactic diffuse cloud spectra are shown 
for comparison (from \citealt{2011A&A...533A.129V}). The normalised spectra are offset vertically 
for clarity and the Galatic spectra were shifted in velocity to match the observed DIBs towards SN\,2008fp.
See the text for more details.}
\label{fig:sn2008fp_dibs}
\end{figure}

However, a recent survey by \citet{2011A&A...533A.129V} showed that the correlation of the 5780 DIB with reddening is different 
for $\sigma$ and $\zeta$-type sightlines\footnote{$\sigma$ and $\zeta$-type diffuse clouds refer to the archetypical Galactic 
lines-of-sight $\sigma$\,Sco and $\zeta$\,Oph. $\sigma$\,Sco probe a diffuse environment exposed to a strong radiation field 
with low molecular content, while $\zeta$\,Oph probes a denser cloud protected from the impinging interstellar radition. 
For a discussion on the two types of clouds and how this connects to molecular content, the ISRF strength and DIB (ratio) we 
refer to \citet{2011A&A...533A.129V}.}. For a $\zeta$-type line-of-sight the 5797~\AA\ DIB is deeper than the 5780 DIB. 
In addition, for SN\,2008fp the 5797/5780 strength ratio is 0.74. Such high values are generally only observed toward dense ($\zeta$-type) 
Galactic sightlines.
We note that the reduced strength of 
the `broad' 5780 and 6283 DIBs with respect to those of the `narrow' DIBs at 5797, 5849 and 6196~\AA\ is typical, thus expected, 
for dense clouds in which the production of the carrier of the former family is apparently less efficient. Figure~\ref{fig:sn2008fp_dibs} 
shows the 5780 and 5797~\AA\ DIBs together with those towards two Galactic sightlines that are typical of a diffuse cloud exposed 
to a strong UV field ($\sigma$-type) and a denser (molecular) cloud shielded from the interstellar UV radiation field ($\zeta$-type).
Scaling the 5780-5797 DIB spectrum observed toward Galactic $\zeta$ type HD\,149757 (\Ebv = 0.32~mag) by about 40\% provides a good 
fit to the two diffuse bands, thus implying that this dense cloud has \Ebv $\approx$ 0.45~mag (Figs.~\ref{fig:sn2008fp_dibs} 
and~\ref{fig:sn2008fp_ISlines}). 
For $R_\mathrm{V} \leq 2$ (Sect.~\ref{discussion:pol}) this corresponds to $A_\mathrm{V} \leq 0.9$~mag.

The narrow profiles of the observed di-atomic molecules and their relative high-abundance together with the inferred high
molecular hydrogen fraction and intermediate visual extinction of 0.9~mag indicates that the line-of-sight towards SN\,2008fp
probes a translucent cloud (\citealt{2006ARA&A..44..367S}).

For Galactic sightlines with high abundances of C$_2$ several DIBs are also observed to be stronger with respect to the other
DIBs. The strongest C$_2$-DIB is the $\lambda$4963 band, although the 4734 and 5769~\AA\ DIBs shows the tightest correlation
with N(C$_2$)/\Ebv\ (\citealt{2003ApJ...584..339T}). Unfortunately, the 4963 and 4737~\AA\ DIBs (at the host galaxy velocity)
are not included in our spectral range. The 5769~\AA\ DIB was not detected (its enhanced EW/\Ebv\ is only 18~m\AA).

We note also that the C$_2$ near-infrared band ($J$ levels from 0 to 8) has recently been detected in the Small Magellanic Cloud, 
thus for the first time beyond the Milky Way, for the peculiar line-of-sight towards Sk\,143 (\citealt{2013MNRAS.428.1107W}). 
The detection of C$_2$ and C$_3$ in the SN\,2008fp host galaxy constitutes the first direct detection of these 
molecules beyond the Local Group. These observations illustrate that optical high-resolution spectroscopy of SNe offers an unique 
opportunity to probe in detail the physics and chemistry of the diffuse cold medium in distant galaxies. In particular, C$_2$ is 
a sensitive probe to local density and temperature, but requires both high-resolution and high S/N in the optical range, which
is currently
only accessible for relatively bright sources even with 10m-class telescopes.

\subsection{Molecular hydrogen fraction}

In this section we attempt to derive an estimate of the molecular hydrogen fraction of the extra-galacitc translucent cloud in the line-of-sight toward SN\,2008fp.
First, the \ion{H}{i} column density can be estimated from the 5780 DIB strength: log(N(\ion{H}{i}) = (19.00$\pm$0.08) + (0.94$\pm$0.04) log(W(5780)) 
(\citealt{2011ApJ...727...33F}), which gives for SN\,2008fp N(\ion{H}{i}) = $6.2^{+3.3}_{-2.1} \times\ 10^{20}$~cm$^{-2}$. 
Then, for Galactic sightlines, the CH column density shows a strong linear relation with the H$_2$ column density: 
N(CH)/N(H$_2$) = $3.5^{+2.1}_{-1.4}$ $\times$ 10$^{-8}$ (see e.g. \citealt{1982ApJ...257..125F}, \citealt{2008ApJ...687.1075S}). 
Thus, N(H$_2$)~=~$7.2^{4.8}_{-2.7} \times 10^{20}$~cm$^{-2}$. 
Now we obtain $f_{H_2} = 2 N(H_2) / ( N(\ion{H}{i}) + 2 N(H_2) )$ = $0.70^{+0.15}_{-0.20}$. 
This confirms our earlier conclusion that the line-of-sight towards SN\,2008fp probes a dense, $\zeta$-type cloud in the host galaxy.

\subsection{Carbon chemistry and abundance}

The ratio CH$^+$/CH scales with the strength of the interstellar radiation field (I$_\mathrm{UV}$), 
density ($n_H$) and molecular hydrogen fraction ($f_{H_2}$) as derived by e.g. \citet{2006ApJS..165..138W}
and \citet{2006ApJ...649..788R}: N(CH$^+$)/N(CH)~$\propto$~I$_\mathrm{UV}$/($n_H~f_{H_2}$).
The measured ratio, N(CH$^+$)/N(CH) = 0.6$^{+1.7}_{-0.44}$, is similar to that observed toward SN\,2006X (\citealt{2008A&A...485L...9C}). 
It is indicative of inefficient CH$^+$ production and thus a quiescent ISM (e.g. \citealt{1989MNRAS.241..575C}, \citealt{2011A&A...533A.129V}). 
However, since the (average) density and molecular fraction are high (typical for translucent clouds), the ambient
interstellar UV radiation field can be relatively strong despite the low ratio CH$^+$/CH.

The observed ratio N(C$_2$)/N(C$_3$) is 12~$\pm$~8. This ratio is, within the uncertainties, similar to 
mean ratio of 28.5~$\pm$~6.0 (ranging from 7.6 to 68.1) for a sample of 13 Galactic sightlines (\citealt{2003ApJ...595..235A}).
Note also that the observed total column densities for C$_2$ towards SN\,2008fp is similar to that measured for 
the Galactic sightlines HD\,204827, which has the highest measured C$_2$ and C$_3$ columns density in the survey 
by \citet{2003ApJ...595..235A}. The observed C$_3$ column density is about 50\% higher towards SN\,2008fp,
which results in a lower ratio than the ratio of 38.3 for HD\,204827.

Adopting the values above, the host galaxy ISM displays a high carbon fraction, 
$f$(C$_2$) = N(C$_2$)/N(H$_\mathrm{total}$) = 9.7 $\times$ 10$^{-8}$, which is 
a factor two higher than the Galactic fraction, $f$(C$_2$)$_\mathrm{MW}$ = 3 -- 6 $\times\ 10^{-8}$, found by 
Van Dishoeck \& Black (1989) for diffuse clouds ($A_\mathrm{V} < 5$~mag), 
and a factor six higher than values found by \citet{1999A&A...351..657G}, 
$f$(C$_2$)$_\mathrm{MW}$ = 0.5 -- 1.5 $\times\ 10^{-8}$.

\subsection{Metal depletion}

Taking the tentative detection of \ion{Ti}{ii} at face value we obtain, together with N(H$_\mathrm{total}$), the gas phase interstellar abundance of titanium. 
Since the ionisation potential of \ion{Ti}{ii} is nearly coincident to that of \ion{H}{i}, \ion{Ti}{ii} is the 
dominant ion in \ion{H}{i} regions (see e.g. \citealt{2010MNRAS.404.1321W}). Most of the titanium in the Galactic ISM is 
depleted into dust grains with less than 1 percent of the total titanium in the gas phase (e.g. \citealt{2009ApJ...700.1299J}).
Like \ion{Ca}{ii}, \ion{Ti}{ii} traces warmer, more smoothly distributed neutral component of the ISM 
(\citealt{1994ApJ...424..748C}). 
The ratio N(\ion{Ti}{ii})/N(\ion{Ca}{ii}) is fairly constant, with a value of 0.3 -- 0.4 for Galactic environments
(\citealt{1983ApJ...272..509A}, \citealt{2006MNRAS.367.1478H}), but is on average higher, at about 0.9, for the SMC and LMC.
\citet{2010MNRAS.404.1321W} showed that there is no significant saturation for the \ion{Ti}{ii} 3383.768~\AA\ line up to 100~m\AA.
For SN\,2008fp the observed ratio of N(\ion{Ti}{ii})/N(\ion{Ca}{ii}) is 0.05, and the relative depletion Ti/H is -9.2 dex, compared
with -7.1 dex for the solar neighbourhood.
Taken at face value, this would imply that metals appear to be heavily depleted into dust grains in the dense extra-galactic 
line-of-sight towards SN\,2008fp. However, given the uncertainties in the different column densities and abundances, we caution
against drawing any firm conclusions.

\citet{1991ApJ...381L..17C} found that N(\ion{Ca}{i})/N(\ion{K}{i}) $\sim\ \delta$(Ca), where the calcium depletion $\delta$(Ca) is $\propto n_H^3$.
For SN\,2008fp the measured ratio of $0.214^{1.74}_{-0.16}$ is low, which gives $\delta$(Ca) $\sim$ 0.017 -- 0.028 
(or log $\delta$(Ca) $\sim$ -1.6 to -1.8), depending on whether photoionisation equillibrium or charge exchange processes are assumed. 
In Galactic environments low ratios are seen for clouds with molecular hydrogen fractions $> 0.1$.

\subsection{Dust polarisation properties}\label{discussion:pol}

The polarisation wavelength dependency shown by SN\,2008fp is reminiscent of what was found in SN\,2006X (\citealt{2009A&A...508..229P}). 
This is illustrated in Fig.~\ref{fig:isp} where, for presentation, the dominant axis component was scaled by a factor 3.3 to
match the data-sets of the two SNe at $\sim$4000~\AA.  Bearing in mind that in SN\,2008fp the intrinsic SN signatures are stronger, the
resemblance is significant.  In both cases the polarisation grows at shorter wavelengths and, if anything, the maximum is attained
below 4000~\AA. As described in \citet{2009A&A...508..229P}, the polarisation wavelength dependency of SN\,2006X can be reproduced by
a Serkowski law with $\lambda_\mathrm{max}$ = 3500~$\pm$~100 \AA\/ and $K = 1.3~\pm~0.1$ (see Fig.~\ref{fig:isp}, short-dashed line). The value of
$K$ is higher than $K = 1.15$ derived by \citet{1975ApJ...196..261S}) and disagrees with the wavelength dependency derived by
\citet{1992ApJ...386..562W}, $K = 0.01 + 1.66 \lambda_{max}$ (where $\lambda_\mathrm{max}$ is expressed in $\mu$m), yielding $K$ = 0.60 for
$\lambda_{max}$ = 0.35~$\mu$m. The corresponding polarisation law is traced in Fig.~\ref{fig:isp} (dotted line), which clearly
illustrates the poor match to the data. 

\citet{1987MNRAS.227P...1H} pointed out that $\lambda_{max}$ for SN\,1986G is lower than typical MW values, 
while it is closer to what is found in M31. 
Assuming that the dust composition is similar, they concluded that the grain sizes must be about 20\% smaller in Cen\,A than in 
the MW and postulated that this should have some consequences on the exctinction law. 
Using the empirical relation derived by \citet{1975ApJ...196..261S} between the selective-to-total absorption $R_\mathrm{V}$ and 
$\lambda_{max}$, $R_\mathrm{V} = (5.6 \pm 0.3) \lambda_{max}$, they derived $R_\mathrm{V} = 2.4~\pm~0.13$ for SN\,1986G. 
This value agrees remarkably well with the photometric derivation: $R_\mathrm{V} = 2.6 \pm 0.2$ (\citealt{2013ApJ...779...38P}).

The cases of SN\,2006X and SN\,2008fp appear to be even more extreme. 
As pointed out in \citet{2009A&A...508..229P}, the lack of a maximum polarisation in SN\,2006X is a convincing argument in
favour of an exceptionally small $R_\mathrm{V}$, giving independent support to the photometric derivations: 1.48~$\pm$~0.06 
(\citealt{2008CBET.1509....1W}), 1.31$^{+0.08}_{-0.10}$ (\citealt{2013ApJ...779...38P})). 
Given the spectropolarimetric data presented here, we argue that this plausibly applies to SN\,2008fp as well. 
Using $\lambda_\mathrm{max}~\leq~3500$~\AA\ in the above relation gives $R_\mathrm{V}~\leq~2$, consistent with
$R_\mathrm{V} = 1.20^{+0.26}_{-0.14}$ derived from the light curve (\citealt{2013ApJ...779...38P}).
Along the lines proposed by \citet{1987MNRAS.227P...1H}, we conclude that the dust grains composition and/or size 
must be substantially different from those typical of the MW (see also \citealt{1992dge..book.....W}).

One possible problem with the ISP determination for SN~2008fp is the unaccounted MW contribution. 
The Galactic foreground reddening is \Ebv\ = 0.2 (\citealt{1998ApJ...500..525S}). 
The host galaxy photometric reddening  estimate is \Ebv\ = 0.6~$\pm$~0.1, 
implying that the Galactic reddening contribution is about one forth of the total. 
In the above discussion we implicitly assumed that the MW contribution is not significant. 
This is supported by the position angle alignment with the host galaxy spiral pattern. 
However, we cannot exclude a contribution to the total polarisation of the order of a several 0.1\%, depending on the angle 
between the host and Galactic polarisation angles.

Given the findings of high-resolution spectroscopy, we conclude that the bulk of reddening, and therefore of polarisation, 
occurs within the molecular material. Whether or not this is related to the progenitors cannot be established based on the data presented here. 
However, given the projected positions of the two events in their hosts, the most natural explanation is that the intervening gas belongs 
to the host galaxy ISM. Irrespective of the location of this material, it remains to be clarified why in these two objects the dust displays properties that deviate so extremely from the typical MW mixture. The question as to how this is related to the growing evidence about 
Type~Ia SN showing systematically small $R_\mathrm{V}$ has to be addressed using statistically significant samples that now start to be 
available (Zelaya et al. in preparation).

\subsection{Cosmic radiation background}\label{sec:CN}

The relative line strengths of the R$_0$ and R$_1$ absorption lines can be used to derive the number of CN molecules in 
the ground- and first-excited rotational states, with which one can determine the excitation temperature of the $J = 0 - 1$ 
transition at 2.64~mm (\citealt{1972ARA&A..10..305T}, \citealt{2011ApJ...728...36R}, \citealt{2012MNRAS.421.1325L}). 
In the absence of local excitation, $T_{01}$ = $(1+z)$\,T$_\mathrm{CMB}$ = 2.7416 $\pm$ 0.003~K 
(T$_\mathrm{CMB-COBE}$ = 2.72548 $\pm$ 0.00057~K; \citealt{1999ApJ...512..511M}; \citealt{2009ApJ...707..916F}) 
and thus N(R$_0$)/N(R$_1$) = 2.463. 
Note that \citet{1985ApJ...297..119M} showed that local collisional excitation of CN by electrons increases $T_{01}$ by only 0.2~K even 
for a high electron density, $n_e = 0.5$~cm$^{-3}$.

Since both R$_1$ and P$_1$ arise from the first excited rotational state of CN, we can use both features, together with R$_0$, to derive $T_{01}$. 
Here we assumed that the observed profile is a composite of at least two individual velocity components with $b \sim 3.4$~\kms, 
derived from decomposing the observed line profile into two components (separated by $\sim$9.6~\kms) of equal intrinsic FWHM of 5.7~\kms.
With the present data we cannot exclude the presence of unresolved narrow ($b \approx 1$~\kms) velocity components such as 
those revealed by high-resolution spectroscopic studies of Galactic translucent clouds (e.g. \citealt{1985ApJ...297..119M}; 
\citealt{1991A&A...251..625G}; \citealt{2005ApJ...633..986P}; \citealt{2008MNRAS.390.1733S}).

Next, we applied the curve-of-growth correction to the strongest velocity component following the method described by 
\citet{1985ApJ...297..119M} (the measurements of the weak redshifted component are of insufficient quality to derive useful quantities).
The results are listed in Table~\ref{tb:CN}.
The mean excitation temperature derived from the line ratios R$_0$/R$_1$ and R$_0$/P$_1$ is $T_{01}$ = $2.9 \pm 0.4$~K for the first epoch.
In addition, we obtain N(R$_0$)/N(R$_1$) = 2.1.
Spectra taken at the second and third epoch are consistent with the first epoch result, but give larger uncertainties. 
We note that although the absolute excitation temperature is sensitive to the adopted $b,$ the difference in derived temperatures is 
much less sensitive. Reducing $b$ to 3.0~\kms\  reduces the temperature to $T_{01}$ = $2.8$~K.
Within the uncertainties on the line measurements there is no evidence for significant local excitation of CN in addition to the excitation
provided by the cosmic microwave background, which is the main reservoir of 2.64~mm photons.

\begin{table}
\caption{Measured equivalent widths and curve-of-growth corrected column densities for CN(0,0) band at each epoch. 
Two velocity components with identical width of 0.11~\AA\ are assumed to contribute to the total absorption profile.
This corresponds to an intrinsic FWHM of 5.7~\kms\ (for resolution of 6.3~\kms) and hence a $b$ of 3.4~\kms.
The equivalent width and column density are given for the total line profile and the strongest velocity component, respectively.
The average CN excitation temperature derived from the R$_0$/R$_1$ and R$_0$/P$_1$ ratios and the 
R$_0$/R$_1$ column density ratios are given for the strongest velocity component only.
}
\label{tb:CN}
\resizebox{\columnwidth}{!}{
\begin{tabular}{llllllll}\hline\hline
 Epoch  &        & \multicolumn{3}{c}{W (m\AA)}               & log N(R$_0$)   & N(R$_0$)/N(R$_1$)   & $T_{01}$    \\ \cline{3-5}
 (days) &        & R$_0$        & R$_1$         & P$_1$       & (cm$^{-2}$)    &                     & (K)         \\ \hline

 $+6$   &$\Sigma$& 87$\pm$6     & 36$\pm$4      & 18$\pm$4    &      13.44$\pm$0.06 &                        &             \\
        & 1      & 70$\pm$5     & 29$\pm$3      & 15$\pm$3    & 13.37$\pm$0.04 & 2.14                        & 2.9$\pm$0.4 \\
        
 $+11$  &$\Sigma$& 79$\pm$6     & 34$\pm$4      & 17$\pm$4    & 13.37$\pm$0.06 &                             &             \\
        & 1      & 63$\pm$4     & 27$\pm$3      & 14$\pm$3    & 13.30$\pm$0.04 & 1.95                        & 3.0$\pm$0.5 \\
        
$+17$   &$\Sigma$& 82$\pm$8     & 36$\pm$7      & 14$\pm$7    & 13.40$\pm$0.08 &                             &             \\ 
        & 1      & 66$\pm$6     & 29$\pm$6      & 11$\pm$5    & 13.33$\pm$0.06 & 1.95                        & 2.8$\pm$0.6 \\  
\hline
\end{tabular} 
}
\end{table}

\subsection{Line variability}

The UVES data span from about 2 days before maximum to 30 after maximum, which translates into a time-coverage of 17 to
50 days after the estimated explosion (assuming the canonical 19-day rise-time of Type Ia SN). In this time span the
photosphere expanded approximately from 120~AU to 240~AU, that
is, doubling its size.  In the case of a patchy ISM with a
fractal structure (\citealt{2010A&A...514A..78P}), the peak-to-peak variations  in this time range are always smaller than a few tens
of m\AA, with the rms variations being of the order of a few m\AA. Therefore, the lack of larger variations, for example the
CN line measurements at phases 1 and 2, excludes small, isolated clouds (or dense compact cloudlets) with sizes similar
to  the SN photospheric radius (100 -- 200~AU). 
A detailed discussion of the (lack of) variability in column denstities of the atomic \ion{Na}{i} and \ion{Ca}{ii} lines 
will be presented elsewhere as part of a larger survey studying the variability in atomic absorption profiles towards Type~Ia supernovae.

\subsection{Probing extra-galactic molecular clouds: SN\,2006X and SN\,2008fp}\label{sec:2006X}

Excitation analyses of the UV/blue CN(0,0) band in SNe host galaxies offer a first step towards an {\it in situ} probe of the CMB
temperature history, as suggested by {\citet{1980ARA&A..18..489W}}. For now this has been possible for relatively nearby SNe\,2006X
and~2008fp, which have redshifts of $z = 0.005240$ and $z = 0.005664$, respectively. We note that for SN\,2006X the CN excitation
temperatures, $T_\mathrm{ex} = 3.0\pm0.2$~K, is also fully consistent with the CMB (\citealt{2007Sci...317..924P}). 

For SN\,2006X the \Ebv\ was estimated to be around 1.2 to 1.4~mag (see e.g. \citealt{2009A&A...508..229P}). 
It appears that SN\,2008fp suffers a lower extinction of \Ebv\ $\approx$0.45~mag.
SN\,2008fp is located at a projected distance of about 2.2~kpc from the host galaxy nucleus, and from its
axial ratio one deduces an inclination of about 55 degrees. Its relatively high extinction furthermore suggests that it is placed at a low galactic
latitude or behind the disk of the host galaxy. This is similar to the position of SN\,2006X in its host galaxy. 

\begin{figure}[t!]
\centering
\includegraphics[angle=0,width=0.9\columnwidth,clip]{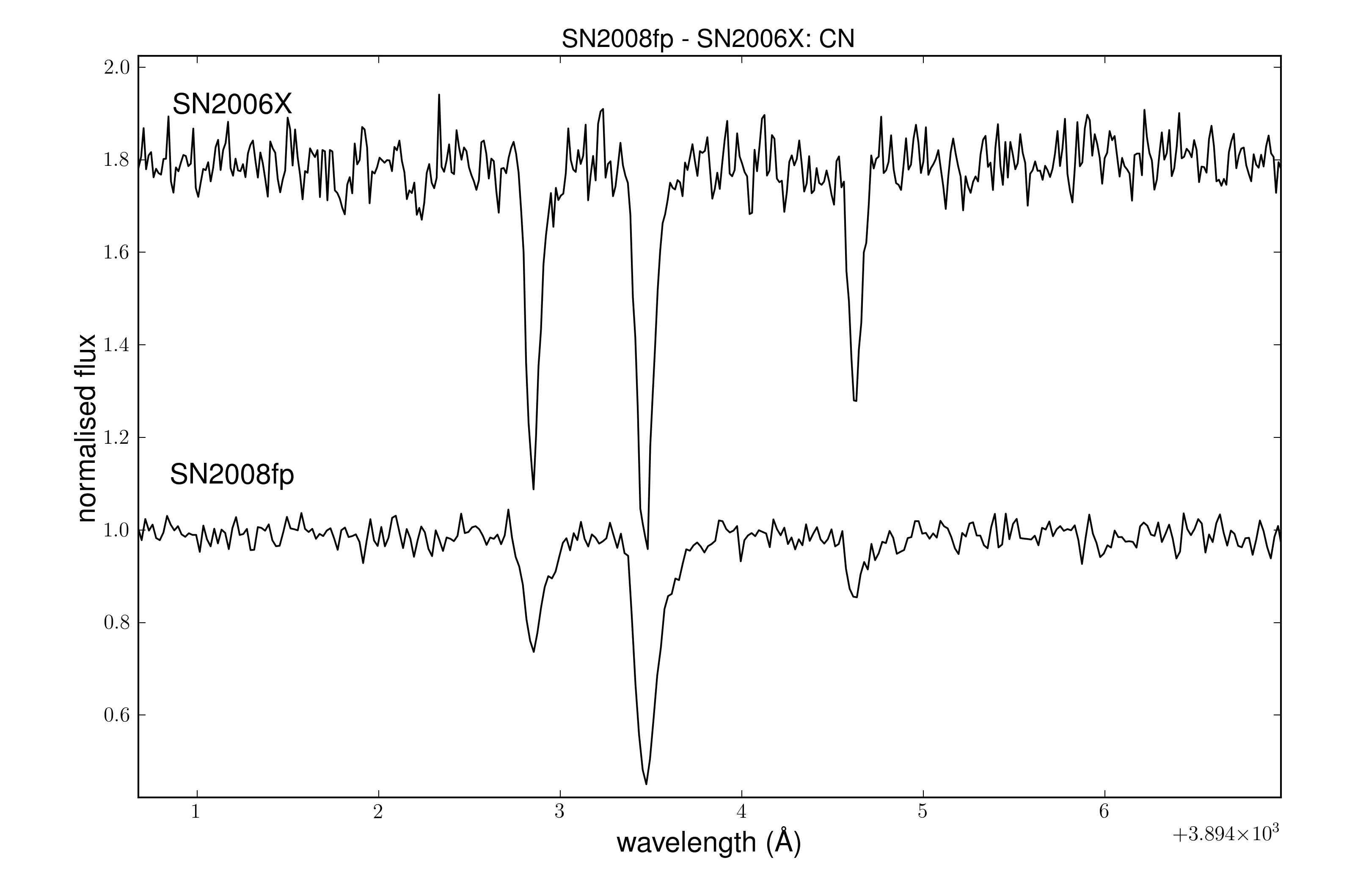}
\includegraphics[angle=0,width=0.9\columnwidth,clip]{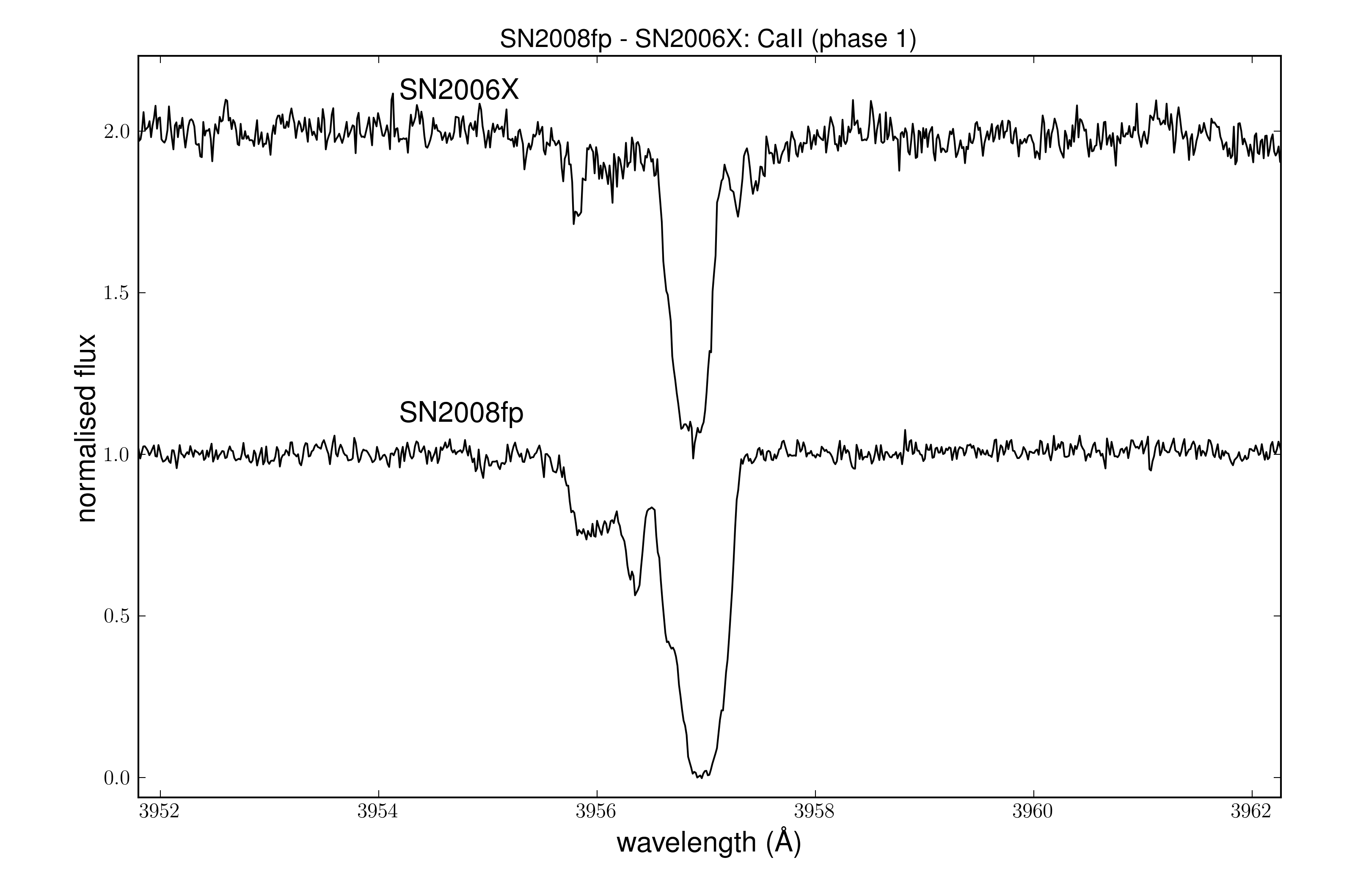}
\caption{CN and \ion{Ca}{ii} line profiles for both SN\,2008fp and SN\,2006X are shown, both at 
2 days before maximum light. The unresolved CN lines in SN\,2006X do not reveal the red wing seen for SN\,2008fp.}
\label{fig:sn2008fp_sn2006X} 
\end{figure}

As discussed above, the extinction measured towards both SNe, in total intensity and
polarised light, is primarily due to the ISM and not any CSM. Thus a direct comparison
of the CN and \ion{Ca}{ii} line profiles presents itself rather naturally
(Fig.~\ref{fig:sn2008fp_sn2006X}).  Although similar, the small difference in line
profile of the individual transitions further substantiates that for SN\,2008fp we
observe two components in CN, instead of it having a higher $b$-value. The observed
similarity with SN\,2006X in terms of CN and extinction, but lack of a variable sodium
component, clearly shows that high extinction does not necessarily imply time variations,
as noted already by \citet{2009ApJ...702.1157S}. At this time we cannot pinpoint
the origin of the unusually strong CN lines towards these two SNe, although it appears to
be related  somehow to a specific subset of Type\,Ia SNe and/or their host galaxies.
Another recent example is SN2009ig, which also shows strong CN absorption lines relative
to the line-of-sight visual extinction (\citealt{2014IAUS..297..106C}).

At face value, the ratio between the $E_{(B-V)}$ of SN\,2006X and SN\,2008fp is $\approx$2.7--3.1, close to the polarisation ratio of 3.3, 
implying that SN\,2008fp has a polarisation that is 10--20\% lower than one would expect if it behaved like 2006X. 
This can be interpreted as a slightly lower efficiency of grain alignment in the intervening material of SN\,2008fp, 
which suffers a reddening higher than its polarisation would indicate, although the effect is minimal. 
One possibility is that this is related to the anomalous ratio between the \ion{Na}{i}\,D equivalent width and reddening 
(\citealt{2013ApJ...779...38P}).

\section{Conclusion}\label{sec:conclusion}

We presented deep high-resolution optical spectra of the Type\,Ia supernovae 2008fp. We focused on the
analysis and interpretation of these spectra in the context of understanding the physical properties of interstellar 
molecular clouds in extra-galactic environments. This analysis of the line-of-sight towards SN\,2008fp revealed the following:

\begin{itemize}

\item In addition to the main atomic species (\ion{Na}{i}, \ion{Ca}{ii}), many weaker lines are detected:
\ion{Fe}{i}, \ion{Ca}{i}, \ion{Ti}{ii}, CH, CH$^+$, CN, as well as diffuse interstellar bands at 5780, 5797, 5849, 6196, and 6283~\AA.

\item The 5780 and 5797~\AA\ DIBs indicate that the $\zeta$-type cloud component at 1770~\kms\ can be considered a translucent cloud 
sufficiently shielded from interstellar UV radiation (from young stars in the host galaxy) to ensure a rich molecular chemistry.
From a comparision with Galactic $\sigma$ and $\zeta$-type DIB spectra we inferred \Ebv = 0.45~mag for the total extra-galactic 
reddening towards SN\,2008fp. This fully agrees with $A_\mathrm{V}$ and $R_\mathrm{V}$ derived from modelling the light curve.

\item The molecular hydrogen fraction of 0.7$^{0.15}_{-0.2}$ corresponds to that of translucent clouds.

\item The C$_2$ (2,0) band is detected beyond the Local Group, in the SN\,2008fp host galaxy.
The rotational line analysis yielded T$_\mathrm{rot}~\approx$~30~K and $n_C~\approx$~250~cm$^{-3}$.
The C$_2$/H$_\mathrm{total}$ fraction is a factor three or more higher than typically found for Galactic diffuse clouds.

\item We presented a column density of N(C$_3$) = 1.7 $\pm$ 0.5 $\times$ 10$^{13}$~cm$^{-2}$ for extra-galactic C$_3$ towards SN\,2008fp, 
which constitutes the first detection of C$_3$ beyond the Local Group.
The relative abundance of C$_3$ with respect to C$_2$ is similar to that of the Galactic mean.

\item The relative depletion of Ti/H is $\approx$ -9.2~dex, which indicates a higher level of depletion than 
the -7.1~dex depletion found for the solar neighbourhood.

\item From the atomic and molecular line analysis we conclude that the bulk of reddening (dust extinction), 
and therefore of polarisation, occurs within molecular clouds in the SN\,2008fp host galaxy along the 
line of sight. If not physically associated with the SN progenitor, this implies that the reddening and 
polarisation arise primarily in the ISM and not in the CSM. This conclusion is strengthened by the polarisation 
position angle. If the dust were of CS nature, it would produce polarisation by scattering, 
which would turn into a non-null net polarisation only if the dust is distributed asymmetrically.

\item The wavelength dependency of interstellar polarisation in the host of SN\,2008fp clearly differs 
from that displayed by similarly reddened stars in the Galaxy. The observed behaviour is very similar to 
that of SN\,2006X, which also shows a total-to-selective extinction ratio $R_V<2$, indicating a dust 
size/composition significantly different from that typical of the Milky Way.

\item From the CN(0,0) band the CN excitation temperature at 2.64~mm is 2.9$\pm$0.4~K, consistent with that of the
cosmic microwave background, and statistically inconclusive regarding local collisional excitations by electrons.

\item The lack of variability over a 28-day period excludes small, isolated clouds with sizes 
similar to the SN photospheric radius of 100 to 200~AU, supporting the presence of a more patchy - fractal - ISM.

\item High-resolution spectra of Type\,Ia SNe reveal several sightlines -- SN\,2006X, SN\,2008fp, SN\,2009ig -- 
with an inexplicably strong CN B-X (0,0) absorption band. Whether and how this is related to a specific subset of Type\,Ia
SNe or their host galaxies remains an open question.

\end{itemize}

\begin{acknowledgements} 

We thank the referee for a careful reading of the manuscript and constructive insights that improved this paper.
We have benefitted from discussions with Dan Welty on the topic of molecular optical spectroscopy.
This research has made use of NASA's Astrophysics Data System Bibliographic Services, of the NASA/IPAC 
Extragalactic Database (NED) which is operated by the Jet Propulsion Laboratory, California Institute of
Technology, under contract with the National Aeronautics and Space Administration, and of the SIMBAD database, 
operated at CDS, Strasbourg, France.

\end{acknowledgements}

\bibliographystyle{aa}
\bibliography{/lhome/nick/Desktop/ReadingMaterial/Astronomy/Bibtex/bibtex}

\end{document}